\documentclass[11pt]{article}
\usepackage[pagebackref,raiselinks,hyperindex]{hyperref}
 
\usepackage{a4}
\usepackage{latexsym}
\usepackage{amssymb}
\usepackage{exscale}
\usepackage{amsfonts,amsmath}
\usepackage{theorem}

\newcommand{\be}{\begin{equation}}
\newcommand{\ee}{\end{equation}}
\newcommand{\bea}{\begin{eqnarray*}}
\newcommand{\eea}{\end{eqnarray*}}
\newcommand{\beq}{\begin{eqnarray}}
\newcommand{\eeq}{\end{eqnarray}}
\newcommand{\nn}{\nonumber}
\newcommand{\RR}{\mathbb{R}}
\newcommand{\CC}{\mathbb{C}}

\newcommand{\NN}{\mathbb{N}}
\newcommand{\ZZ}{\mathbb{Z}}

\newcommand{\EE}{\, \mathbb{E} \,}
\newcommand{\PP}{\mathbb{P}}

\newcommand{\NO}{| \| }
\newcommand{\ON}{\| | }

\newcommand{\ra}{\Longrightarrow  \;}
\newcommand{\aeq}{\Longleftrightarrow}

\newcommand{\g}{\ZZ^d}
\newcommand{\df}{\frac {\partial \tilde{f}} {\partial \bar{z}}}
\renewcommand{\H}{H_{\omega}}
\newcommand{\h}{H_{\omega,l}}
\renewcommand{\r}{\right}
\renewcommand{\l}{\left}
\newcommand{\diam}{\mathop{\mathrm{diam}}}
\newcommand{\const}{\mathop{\mathrm{const.}}}

\newcommand{\supp}{\mathop{\mathrm{supp}}}
\newcommand{\Id}{{\mathop{\mathrm{Id}}}}
\newcommand{\Tr}{\mathop{\mathrm{Tr}}}

\newtheorem{thm}{Theorem}[section]

\newtheorem{prp}[thm]{Proposition}
\newtheorem{lem}[thm]{Lemma}
\newtheorem{kor}[thm]{Corollary}
\newtheorem{dfn}[thm]{Definition}

\theorembodyfont{\normalfont}
\newtheorem{bem}[thm]{Remark}

\newtheorem{bsp}[thm]{Example}

\newenvironment{pro}{\pagebreak[2]  \mbox{} \\ {\bfseries Proof:} \par \nopagebreak }
            {\par \hspace*{\fill} {\bfseries q.e.d.} \\[1ex]}

\begin{document}

%
%
%
%
\title{
Localization for random perturbations\\ 
of periodic Schr\"odinger operators\\
with regular Floquet eigenvalues%
\footnotetext{
{For somewhat different versions see: 
\href{http://www.ma.utexas.edu/mp_arc-bin/mpa?yn=98-569}{http://www.ma.utexas.edu/mp\_arc-bin/mpa?yn=98-569}
\\ and \href{http://dx.doi.org/10.1007/s00023-002-8621-x}{ Annales Henri Poincar\'e , Volume 3, Issue 2, pp 389-409 (2002)}}.} 
}
\author{{ Ivan Veseli\'{c}}\\[1ex]
{Fakult\"at f\"ur Mathematik,}\\
{Ruhr-Universit\"at Bochum,}\\
{D-44780 Bochum, Germany,}\\
{and SFB 237}\\
{\href{http://www.ruhr-uni-bochum.de/mathphys/ivan/}{http://www.ruhr-uni-bochum.de/mathphys/ivan/}}
}
\date{}
\maketitle
\begin{abstract}
We prove  a localization theorem for  continuous ergodic Schr\"odinger operators $ H_\omega := H_0 + V_\omega $,
where the random potential $ V_\omega $  is a nonnegative Anderson-type perturbation of the periodic operator $ H_0$.
We consider a lower spectral band edge of $ \sigma ( H_0) $,
say $ E= 0 $, at a gap which is preserved  by the perturbation $ V_\omega $.
Assuming that all Floquet eigenvalues of $ H_0$, 
which reach the spectral edge $0$ as a minimum, 
have there a positive definite Hessian,
we conclude that there exists an interval $ I  $ containing $0$
such that $ H_\omega $ has only pure point spectrum in $ I $  for almost all $ \omega $.
\end{abstract}

\section{Introduction and results\label{sec1}}

\subsection*{Localization}

Already in the fifties Anderson \cite{Anderson-58} concluded by physical reasoning
that some random quantum Hamiltonians on a lattice should exhibit
{\it localization} in certain energy regions. That is to say that the
corresponding self-adjoint operator has pure point spectrum in these energy
intervals.

Since then mathematical physicists developed a machinery to prove
rigorously this phenomenon from solid state physics. Most of them used 
the so-called {\it multi scale analysis} (MSA) introduced in a paper by
Fr\"ohlich and Spencer \cite{FroehlichS-83} to prove a weaker form of localization at
low energies for the discrete analogue of the Schr\"odinger operator.
This quite complicated reasoning was streamlined by von Dreifus and Klein
\cite{DreifusK-89}. The underlying lattice structure made the MSA easier to apply to
discrete Hamiltonians but soon adaptations for continuous Schr\"odinger
operators followed \cite{MartinelliH-84,Klopp-1995,CombesH-94b,Klopp-95a}.
We prove in Theorem \ref{physThm} a localization result for energies near internal spectral edges of a periodic
Schr\"odinger operator $H_0$ which is perturbed by an Anderson-type potential $V_\omega$. Unlike \cite{BarbarouxCH-1997,KirschSS-1998a} our results are not restricted to a special disorder regime of the random coupling constants
in $V_\omega$. Instead we assume that the periodic operator $H_0$ has regular Floquet eigenvalues. This behaviour is commonly assumed among physicists. Recent results by Klopp and Ralston indicate that it is generic \cite{KloppR-2000}.

In the remainder of this section we introduce our model, state the main Theorem \ref{physThm} and the technical Proposition \ref{techThm} on which it is based. Section 2 explains how to deduce  Theorem \ref{physThm} from
Proposition \ref{techThm}, in Section 3 we describe the functional calculus with almost analytic functions,
Section 4 contains a comparison result between the integrated density of states on finite cubes and on the whole of $\RR^d$
and the last Section 5 deals with periodic (or more generally quasi-periodic) boundary conditions which are necessary to complete the proof of Proposition \ref{techThm}. Two technical proofs are placed in an appendix.

\subsection*{The model}
On the Hilbert space $ L^2 (\RR^d) $ we consider a self-adjoint operator
$ H :=H_\omega $ made up of a periodic Schr\"odinger operator $H_0$
and a random perturbation $V_\omega$
\be
H_\omega := H_0 + V_\omega \ .
\ee
Here $ H_0 := -\Delta + V_0 $ is the sum of the negative Laplacian and a
$ \ZZ^d$-periodic potential $ V_0 \in L^p_{loc} (\RR^d) $ with
$ p=2 $ if $ d \le 3 $, $ p > 2 $ if $ d = 4 $ and $ p \ge d/2 $ if $ d \ge 5 $.
Such a potential is an infinitesimal perturbation of $ - \Delta $
so the sum is self-adjoint with domain $ D(- \Delta) =W^2_2 (\RR^d) $,
the Sobolev space of $L^2$-functions whose second derivative is also in $L^2$
(cf.~\cite{ReedS-75,ReedS-78}).
The random perturbation is of {\it Anderson type}
\be
V_\omega (x) := \sum_{k \in \ZZ^d} \omega_k \,  u(x -k)  \  ,
\ee
where $ ( \omega_k )_{k \in \ZZ^d} $ is a collection of independent identically
distributed (i.i.d.) random variables, called {\it coupling constants}. Their distribution
has a bounded density with support $ [0, \omega_{\max}] $ for some $ \omega_{\max} >0 $.
The non-negative {\it single site potential} $ u $ has to decay exponentially and have
an uniform
lower bound on some open subset of $ \RR^d $, more precisely
\[
u \ge \delta_1 \chi_\Lambda,  \ \delta_1 > 0
\mbox{ where } \Lambda := \Lambda_s := \{ x \in \RR^d | \ \|x\|_\infty < s/2 \}
, \ s > 0
\]
and
\be
\label{uDecay}
\| \chi_{\Lambda_{1}} u ( \cdot -l) \|_{L^p} \le \delta_2 \, e^{-\delta_3 l}  , \ \delta_2 ,\delta_3 > 0 \ .
\ee
%
%
$H_\omega $ is an ergodic operator and we infer from \cite{Kirsch-89a,CarmonaL-90} or
\cite{PasturF-92} that there exists a set $ \sigma \subset \RR $ such that
$ \sigma = \sigma( H_\omega ) $ for almost all $ \omega \in \Omega $, i.e.
the spectrum of $ H_\omega $ is almost surely non-random.
In the same sense $ \sigma_{ac},\sigma_{sc} $ and $ \sigma_{pp} $ are 
$ \omega$-independent subsets of the real line.

Under some mild assumptions the periodic background operator $ H_0 $ has a
spectrum with {\it band structure}, i.e. $ \sigma( H_0 ) =
\bigcup_{n \in \NN} [E^-_n,E^+_n]  \, , \  E^-_1 \le E^+_1 \le E^-_2 \le \ldots $,
 where  for some $ n $ we have open spectral gaps, i.e.~$ E^+_n < E^-_{n+1} $ (cf.~ \cite{Eastham-73,Sjoestrand-91,ReedS-78}).
We assume that there exist positive numbers $ a,b $ and $ b'$ with
\bea
[0,a] \subset \sigma(H_0), \ [-b,0[ \subset \rho(H_0)  
 \mbox{ and } [-b',0[ \subset \rho(H_\omega) \, .
\eea
Since $0$  is in the support of the density of $\omega_0$ it follows that $0 \in \sigma( H_\omega)$.
In this case we say that $ 0 $ is a {\it lower band edge} of the periodic
operator, which is preserved by the positive random perturbation $ V_\omega $.

$H_0 $ can be decomposed into a {\it direct integral} via an unitary
transformation $ U $ (cf.~\cite{Sjoestrand-91,ReedS-78})
\be
\label{dirint}
U H_0 \, U^* = \int^\bigoplus_{[-\pi,\pi]^d} H_0|^\theta_{\Lambda_{1}} \, d\theta
\ .
\ee
Here $ H_0 |^\theta_{\Lambda_{1}}  $ is the same formal differential expression as $ H_0 $ 
acting on functions $ f \in W^2_2
( \Lambda_{1}) $ with {\it $\theta$-boundary conditions}, i.e. for all
$j= 1,\ldots, d $ we have a phase shift in the corresponding direction:
$ f(x + e_j) = e^{i\theta_j} f(x)$ where $ x_j = -1/2 $.
It is an operator with discrete spectrum, which consists of the so-called
{\it Floquet eigenvalues}
\[
E_1(\theta) \le \ldots \le E_n(\theta) \le \ldots \hspace{2em} n \in \NN \ .
\]
These are Lipschitz-continuous on $ [- \pi , \pi ]^d $.
In fact they "generate" the bands of the spectrum of $ H_0$
\bea
\sigma ( H_0) = \bigcup_{n \in \NN} \bigcup_{\theta \in [ - \pi ,\pi ]^d} E_n (\theta) \ .
\eea
There is a finite set of indices $ \cal{N} \subset \NN $  (cf.~\cite{Sjoestrand-91})
such that
\[
E_n(\theta) =0 \mbox{ for some } \theta \in  [- \pi , \pi ]^d \ra n \in \cal{N}
\ .
\]
Since $ 0 $ is a lower band edge of $ \sigma(H_\omega) $, $ E_n(\theta) =0 $ has to
be a minimum of $ E_n (\cdot ) $. If for all $ n \in \cal{N} $, $ E_n (\cdot ) $
has only quadratic minima at $0$ (i.e. the Hessian of $ E_n (\cdot ) $ at any minimum
with value $0$
is positive definite) we say that $  H_0 $ has {\it regular Floquet eigenvalues
at $ 0 $}. 

\subsection*{Results}
Our result on localization  at an lower internal spectral band edge is the following 
\begin{thm}
\label{physThm}
If $ H_0 $ has regular Floquet eigenvalues at $ 0 $ and $ H_\omega $ is
constructed as above, then there exists a number $ E_0 > 0 $ such that
\[
[0,E_0] \subset \sigma_{pp} (H_\omega), \  [0,E_0] \cap \sigma_{c}( H_\omega ) = \emptyset   \ .
\]
\end{thm}
The proof of the theorem is based on the following proposition.
\begin{prp}
\label{techThm}
Assume that $ H_0 $ has regular Floquet eigenvalues at $ 0 $ and $ H_\omega $ is constructed
as above. 
Then for all $ q > 0 $ and $ \alpha \in ] 0,1[ $ there exists a $ l_0 := l_0(q,\alpha) \in \NN $  
such that for all $ l \ge l_0 $ we have
\[
\PP \{ \omega | \ \sigma( H_\omega|^{per}_{\Lambda_{l}} )
\cap [0,l^{-\alpha}[ \neq \emptyset \: \} \le  l^{-q} .
\]
Here the index ''$per$'' denotes periodic boundary conditions on the cube $\Lambda_{l}$.
\end{prp}
The statement of Proposition \ref{techThm} remains true if we replace the periodic boundary conditions by
general $\theta$-boundary conditions with $\theta \in [ \frac{- \pi}{2l +1}, \frac{ \pi}{2l +1}  ]^d $, cf.~(\ref{dirint}) and (\ref{BriZ}).
The proof of the proposition is given in sections \ref{sec3} to \ref{sec5}.
It uses the existence of {\it Lifshitz-tails} of the {\it integrated density of states} (IDS) of the ergodic operator $ H_\omega $ if $ H_0 $ has regular Floquet eigenvalues, which was proved by Klopp in \cite{Klopp-1999}, who also noted that his result could be used for a localization proof.

Theorem \ref{physThm} is proved using the MSA. Since this technique is well understood by now \cite{CombesH-94b,KirschSS-1998a,Stollmann-2001} we only sketch it to show how Proposition \ref{techThm}, which is the main technical novelty of this paper, enters. This is done in Section 2, where also a discussion of
previous results can be found.


\begin{bem}
At any lower band edge  one can prove localization under
the analogous assumptions. Here $ E = 0 $ was chosen only for notational
simplicity. If the Anderson-type perturbation $ V_\omega $ is negative
our theorem can be used to establish localization on any upper band edge
with regular Floquet eigenvalues.

If the underlying $\ZZ^d$ is replaced by  some other Euclidean lattice
\[
\Gamma := \{ \gamma \in \RR^d | \ \gamma = \sum^d_{j=1} \beta_j a_j, \beta \in
\ZZ^d \}  \ ,
\]
where $ \{ a_j \}^d_{j = 1} $ is a basis of $ \RR^d $, the same theorem and
proposition are valid by a simple modification of the proofs.

In any case we will use the maximum norm when considering lattice points $ k $ or $ \gamma $ in $ \ZZ^d $ or $ \Gamma $, i.e. 
$ |\gamma| := \| \gamma \|_\infty := \max \{ | \gamma_j | , \ j=1, \ldots,d \} $, 
where $ ( \gamma_1, \ldots , \gamma_d ) \in \RR^d $ 
are the components of $ \gamma $.

An inspection of our proofs and the papers \cite{Klopp-1999,Klopp-2000} and \cite{KirschSS-1998a,Zenk-1999} shows 
that  Proposition \ref{techThm} and Theorem \ref{physThm}  extend to single site potentials $u$ with sufficiently fast polynomial decay  (in $L^p$-sense), cf.~(\ref{MinDecOfu}). 
\end{bem}
\begin{bsp}
Finally we give an example of a periodic operator which has only regular
Floquet eigenvalues at all band edges. Thus we know that our condition in the
above theorem is fulfilled and we can prove localization at any lower band edge.
Let $ V_0 $ satisfy the conditions posed above on the periodic potential and
let it be a sum of potentials $ V_j $ which are periodic in the $j$th
coordinate direction and constant in all the others; more precisely  
\[
V_0(x) := \sum^d_{j=1} V_j(x_j)
\]
where $ V_j : \RR \to \RR $ is a periodic function and
$ x = (x_1, \ldots ,x_d) \in \RR^d $. Then both $ H_0 $ and
$ H_0|^\theta_{\Lambda_{1}} $ can be decomposed into a direct sum of
one-dimensional operators. For these it is known that all Floquet
eigenvalues are regular \cite{Eastham-73,Klopp-1999}. As the eigenvalues of the direct sum
are just sums of the eigenvalues of the one-dimensional operators
it is clear that the former also have to be regular.
\end{bsp}
\begin{kor}
\label{explThm}
Let the ergodic operator $ H_\omega := - \Delta + V_0 + V_\omega $ be
constructed as above and the periodic potential be decomposable, i.e.
\[
V_0(x) := \sum^d_{j=1} V_j(x_j)           \  .
\]
Let $ E $ be a lower spectral band edge of the periodic operator
$ H_0 := - \Delta + V_0 $ at a spectral gap
which is not closed by the perturbation $ V_\omega $. Then there exists an
interval $ I \ni E $ such that
\[
I \subset \sigma_{pp} ( H_\omega) , \ \sigma_{c}  (H_\omega) \cap I = \emptyset .
\]
\end{kor}
\subsection*{Acknowledgements}
The author would like to thank F.~Klopp for his hospitality at Universit\'{e} Paris 13, 
for stimulating discussions, as well as for many 
detailed explanations  concerning his paper \cite{Klopp-1999}, W.~Kirsch, under whose guidance
this research was undertaken and H.~Najar, K.~Veseli\'{c} and R.~Muno for valuable comments.

%
%

\section{Multi scale analysis and associated ideas\label{sec2}}

In this section we explain how Theorem \ref{physThm} is deduced
from Proposition \ref{techThm} and discuss previous localization results.

An intermediary step in the proof of localization is the establishing of the exponential
decay of the resolvent
\be
\label{ExpDecOnR}
\sup_{ \epsilon \neq 0} \| \chi_{x} R(\epsilon) \chi_y \|_{ {\cal L} ( L^2 (\RR^d))}
\le  const \, e^{ -c |x-y|} 
\mbox{  for almost all } \omega  \ ,
\ee
where $ R := R(\epsilon) :=  ( H_\omega -E -i \epsilon )^{-1}$ is the resolvent of $ H_\omega $
near an energy value $ E $ in the energy interval $ I  \in \RR$ 
for which we want to prove localization.
The $ \chi_x $ and $ \chi_y $  are characteristic functions of unit cubes
centered at $ x $, respectively at $ y $. This bound can be used to rule out absolutely 
continuous spectrum \cite{MartinelliS-85} and is interpreted as absence of diffusion 
\cite{FroehlichS-83,MartinelliH-84} in the energy
region $ I $  if  (\ref{ExpDecOnR}) holds for all $ E \in I $.

It turns out that the finite size resolvent $ R_\Lambda(\epsilon) := 
( H_\omega |_\Lambda - E -i \epsilon )^{-1} $ is easier approachable
than $ R(\epsilon) $ on the whole space.
Here $ H_\omega |_\Lambda $ is the restriction 
of $ H_\omega $ to $ L^2 ( \Lambda ) $ with some appropriate boundary conditions
(b.c.); the use of Dirichlet or periodic b.c.~is most common.
However the operator $ H_\omega |_\Lambda $ is not ergodic and for its resolvent an estimate
like (\ref{ExpDecOnR}) can be expected to hold only with a probability
strictly smaller  than one. 
This is the place where MSA enters. It is an
induction argument over increasing length scales $ l_j $. They are defined recursively
by $ l_{j+1} := [ l_j^\zeta]_{3} $, where $ [ l_j^\zeta]_{3} $ is the greatest multiple of $3$ smaller than
$ l_j^\zeta $. The scaling exponent $ \zeta $ has to be from the interval $ ]1,2[$.
On each scale one considers the box resolvent
$ R_j(\epsilon) := R_{\Lambda_j}(\epsilon) $ and proves  its exponential decay with a
probability which tends to $1$ as $ j \to \infty $. We outline briefly
the ingredients of the MSA as it is given in \cite{CombesH-94b,KirschSS-1998a}
or \cite{CarmonaL-90}. 

First we explain some notation which is used afterwards. Let  $ \delta > 0 $
be a small constant independent of the length scale $ l_j$ and $ \phi_j(x) \in C^2$
a function which is identically equal  to $0$ for $x$ with $ \| x\|_\infty > l_j -\delta$
and identically equal to one for $x$ with $ \| x\|_\infty < l_j - 2 \delta $.
The commutator 
$ W(\phi_j ) := [-\Delta, \phi_j]:= -(\Delta \phi_j) -2  (\nabla \phi_j) \nabla$
is a local operator acting on functions which live on a ring of width $ \delta $
near the boundary of $ \Lambda_j := \Lambda_{l_j} $. We say that a pair 
$ ( \omega,\Lambda_j ) \in \Omega \times {\cal B } ( \RR^d) $ is {\it 
$m$-regular}, if
\be
\label{ExpDecOnLj}
\sup_{\epsilon \neq 0} \| W (\phi_j ) R_j( \epsilon ) \chi_{l_j/3} \|_{\cal L}
\le e^{-ml_j} \ .
\ee
Here $\| \cdot \|_{\cal L}$ is the operator norm on $L^2(\Lambda_j)$ and
$ \chi_{l_j/3} $ the characteristic function of
$ \Lambda_{l_j/3} := \{ y| \, \| y  \|_\infty \le l_j /6 \} $. Thus the
distance of the  supports of $ \nabla \phi_j $ and $ \chi_{l/3} $ is at least
$ l_j/3 -2\delta \ge l_j/4 $.

Let $q_0 > 0$ and $m_0 \ge const \, l_0^{-1/4}$. The starting point of the MSA is the estimate
\bea
\mbox{(H1)} (l_0,m_0, q_0 ) \makebox[12ex]{}
\PP \{ \omega | \, ( \omega , \Lambda_0 ) \mbox{ is $m_0$-regular} \} \ge 1 - l_0^{q_0}
\eea
which serves as the base clause of the induction. The induction step consists in proving
\be 
\label{H1Step}
\mbox{(H1)} (l_j,m_j, q_j ) \ra \mbox{(H1)} (l_{j+1},m_{j+1}, q_{j+1} )
\ee
For the mass of decay $ m_{j+1} $ and the probability exponent
$ q_{j+1} $ on the scale $ l_{j+1} $ the following estimates are valid 
\beq
\lefteqn{\forall \xi > 0  \ \exists c_1, c_2 , c_3 \mbox{  independent of $j$ such that}}
\nn
\\
m_{j+1}  
& \ge &
 m_j \l ( 1 - \frac{ 4 l_j}{ l_{j+1}} \r ) - \frac{c_1}{l_j} -c_2 \frac{\log l_{j+1}}{l_{j+1}}
\label{mj+1} 
\\
\label{qj+1}
l_{j+1}^{q_{j+1}} 
&  \le &
 c_3 \l ( \frac{l_{j+1}}{ l_j} \r)^{2d} l_{j}^{2q_j} + \frac{1}{2} l_{j+1}^{-\xi} \ .
\eeq
For the recursion clause (\ref{H1Step}) a {\it Wegner estimate} \cite{Wegner-81} is needed:
\bea
\mbox{(H2)}  \makebox[12ex]{}
\PP \{ \omega | \, d( \sigma (H_\omega |_\Lambda ), E ) \le \eta \} 
\le C_W  \: \eta | \Lambda |^2
\eea
for all boxes $ \Lambda \subset \RR^d $ and all $ \eta > 0 $,  such that $ [E-\eta,E+\eta] $
is contained in a suitable small
energy interval near the spectral band edge (cf.~Theorem 3.1 in \cite{KirschSS-1998a}).
Here $ | \Lambda | $ stands for the Lebesgue measure of the cube $ \Lambda $.

The deterministic part of the induction step uses the {\it geometric resolvent formula}
\cite{CombesH-94b,HislopS-96} 
\be
\phi_\Lambda ( H_{\Lambda '} - z )^{-1} = ( H_{\Lambda} - z )^{-1} \phi_\Lambda +
( H_{\Lambda} - z )^{-1} W (\phi_\Lambda) ( H_{\Lambda '} - z )^{-1}
\ee
for $ z \in \rho (H_{\Lambda '}) \cap  \rho ( H_\Lambda ) $ and $ \phi_\Lambda \in C^2$ with support in 
$ \Lambda \subset  \Lambda ' $. It gives the estimate
\be
\| \chi_{l/3}(\cdot - x) R_{3l'} (\epsilon)   \chi_{l/3}(\cdot-y) \|_{\cal L}
\le ( 3^d e^{-ml})^{3|x-y| l^{-1} -4 } \| R_{3l'}(\epsilon) \|_{\cal L}
\ee
if no two disjoint non-regular boxes $ \Lambda_l \subset \Lambda_{l'} $ with center in
$ \frac{l}{3} \g \cap \Lambda_{3l'} $ exist for $ \omega $. 
In our case $ l := l_j$ is the length scale on which the exponential decay of the resolvent is already known and $ l' := l_{j+1} $ the scale on which we want to prove it. 
By the estimates (H1),(H2) we have with probability $1-l_{j+1}^{q_{j+1}} $ (bounded by the inequality (\ref{qj+1})) exponential decay on the length scale
$l_{j+1}$ with mass $m_{j+1}$ (bounded as in (\ref{mj+1})).

We stated above the ingredients of the MSA as they are valid if $ u $ is supported in 
$ \Lambda_{1} $. 
If the single site potential is of long range type as in (\ref{uDecay}) one has to use the
adapted MSA from the papers \cite{KirschSS-1998a,Zenk-1999}.

Once the estimate (H1) is established on all length scales $ l_j, j \in \NN $, one infers
an exponential decay estimate for the resolvent on the whole of $ \RR^d$. Afterwards one
uses a 
spectral averaging technique (cf.\cite{CombesH-94b}) based on ideas of Kotani, Simon, Wolf and
Howland to conclude localization \cite{KotaniS-87,SimonW-86,Howland-87a}.
An alternative version of the MSA can be found in the recently published book \cite{Stollmann-2001}.

Recent papers concentrate on proofs for the Wegner estimate and the initial length scale
decay of the resolvent. At the same time adaptations of the MSA for various random
Schr\"odinger operators, as well as Hamiltonians governing the motion in classical physics 
appeared \cite{FigotinK-1996,FigotinK-1997a,CombesHT-1999,Stollmann-1998}.
Recently Najar \cite{Najar-2001} obtained analog results to \cite{Klopp-1999} and the present paper concerning Lifshitz tails and localization for acoustic operators.

 We discuss briefly some results for quantum mechanical Hamiltonians.

In \cite{Klopp-95a} Klopp proved a Wegner lemma for energies at the infimum of the spectrum 
which applies to an Anderson perturbation $ V_\omega $ with single site potentials $u$
that are allowed to change sign, cf.~also \cite{Veselic-2000b,HislopK-2001}. For $V_\omega $ a Gaussian random field a Wegner estimate
was shown in \cite{FischerHLM-1997}. Its main feature is that no underlying lattice structure of
 $ V_\omega $ is needed.  This result allows one to conclude localization  for the 
corresponding Schr\"odinger operator at low energies \cite{FischerLM-2000}. Kirsch, Stollmann and 
Stolz proved in \cite{KirschSS-1998a} (cf.~also \cite{Zenk-1999}) a Wegner estimate with only polynomial decay conditions on the
single site potential $u$ and deduced a localization result for Hamiltonians 
with long range interactions. They require 
\be
\label{MinDecOfu}
|u(x)| \le const \,  (|x|+1 )^{-m}  \mbox{ for some } m > 4d  \ .
\ee

 The resolvent decay estimate (H1) for some initial length scale can be proved 
with semiclassical techniques. Using the Agmon metric one can achieve rigorously
decay bounds with what is called among physicists WKB-method \cite{CombesH-94b,HislopS-96}.
However this reasoning is only applicable for energies near the bottom of the spectrum.

The so-called {\it Combes-Thomas argument} \cite{CombesT-73} allows one to infer the following inequality
\be
\label{CoThBound}
\| \chi_x (H-z)^{-1} \chi_y \|_{\cal L} \le const \,  d(\sigma(H),z)^{-1}  \,
e^{-const \, d(\sigma(H),z) \, |x-y| }
\ee
where $H$ is a self-adjoint Schr\"odinger operator on $ L^2 (\RR^d) $ and $ z \in \rho(H) $.
It was first applied to multiparticle Hamiltonians \cite{CombesT-73}, but it is also useful in
our case, as soon as we get a lower bound on  $ d(\sigma(H_\omega |_\Lambda),z) $.
Thus it is  sufficient to prove an estimate like
\be 
\label{ProbNoSpecInI}
\PP \{ \omega | \ d(\sigma( H_\omega|_{\Lambda_{l}} ,I)  < l^{- \alpha } /2  \: \} \le  l^{-q}
\ee
for some $ \alpha \in ]0, 1/4] $. Such a bound follows immediately from Proposition \ref{techThm} with $ I := [0, \frac{1}{2} \: l^{-\alpha} [ $,
for $l > (2b')^{-1/\alpha}$. Now Inequality (\ref{CoThBound})
implies the initial scale estimate (H1) with $ m_0 \ge const \, l^{-1/4} $ for $l$ large and $E\in I$, cf.~ 
\cite[Lemma 5.5]{KirschSS-1998a}. 
The constant depends on the energy and the potential, but not on $l $ and $m_0$.

Two possibilities were used to deduce (\ref{ProbNoSpecInI}). The first is to assume a special disorder regime, 
more precisely to demand a sufficiently fast decay of the density $g$ of the distribution of $ \omega $
near the endpoints $ 0$ and $ \omega_{\max} $ of $ \supp g$:
\bea
\lefteqn{\exists \tau > d/2 \ : \ \forall \mbox{ small  } \epsilon >0  } & &
\\
& &
\int_0^\epsilon g (s) ds \le \epsilon^\tau , \mbox{ respectively } 
\int_{\omega_{\max} -\epsilon }^{\omega_{\max} } g(s) ds \le \epsilon^\tau
\eea
depending on whether one wants to consider a lower or upper band edge. This approach
was used in \cite{CombesH-94b,KirschSS-1998a}. Its shortcoming is that it excludes quite a few
distributions, e.g. the uniform distribution on $[0,\omega_{\max}]$.

The other way to prove (\ref{ProbNoSpecInI}), which we pursue, is to use the existence 
of Lifshitz tails of the integrated density of states at the edges of the spectrum.
One defines the IDS usually as follows:
\beq
\label{DefIDS}
N(E) & := & \lim_{\Lambda \nearrow \RR^d} N( H_\omega |_\Lambda^D ,E )
\\
& := & 
\lim_{\Lambda \nearrow \RR^d}  |\Lambda|^{-1} 
\# \{ \mbox{ eigenvalues of } H_\omega |_\Lambda^D
\mbox{ below } E \} \ ,
\eeq
i.e. as the limit of the normalized counting function of eigenvalues of a box Hamiltonian. 
Here $ H_\omega |_\Lambda^D $  is the restriction of $\H $ to $ L^2 (\Lambda) $ with Dirichlet b.c.
As $H_\omega |_\Lambda^D $ has compact resolvent and hence discrete spectrum, definition
(\ref{DefIDS}) makes sense.
$ N(E) $ is almost surely $\omega $-independent and the use of Dirichlet b.c.~in its 
definition implies \cite{KirschM-82c}
\be
\label{SupIDS}
N(E) = \sup_{\Lambda \nearrow \RR^d} N( H_\omega |_\Lambda^D ,E ) \ .
\ee
One says that $ N( \cdot ) $ exhibits Lifshitz tails at some spectral edge $ \cal E $ if
\be
\label{LifTaildef}
\lim_{E \to \cal E} \frac{\log |\log | N(E) - N(\cal E) | |}{\log |E -\cal E|}
= -\frac{d}{2}   \  .
\ee
At the infimum of the spectrum, i.e. for $ {\cal E} = \inf \sigma ( \H ) $, (\ref{SupIDS}) 
and (\ref{LifTaildef}) imply
\bea 
\# \{ \mbox{eigenvalues of } H_\omega |_\Lambda^D \mbox{ in } [{\cal E},E] \}
\le | \Lambda | N(E) \le | \Lambda | \exp ( -c E^{-d/4} ) 
\eea
since $N({\cal E} ) = 0$.
This estimate was used in \cite{Klopp-95a} together with a \v{C}ebi\v{s}ev inequality
to prove (H1) at the bottom of the spectrum, see also \cite{MartinelliH-84}.

If one considers an internal band edge $\cal E $, Lifshitz asymptotics are not so easy to exploit 
since (\ref{SupIDS}) cannot be directly used to bound
\bea
| N ( H_\omega |_\Lambda , {\cal E} ) -  N ( H_\omega |_\Lambda ,  E ) |  \  .
\eea
Therefore a comparison technique between $ N( \cdot ) $ and 
$  N ( H_\omega |_\Lambda , \cdot ) $ is needed. In the one-dimensional case Mezincescu \cite{Mezincescu-93}
proved Lifshitz tails  at internal band edges as well as a comparison lemma for the IDS
(Lemma 2, in Section 4). This proof relies on the delicate analysis of Dirichlet eigenfunctions
of $ \H |_\Lambda $ and their roots. The results in \cite{Mezincescu-93} make a localization proof 
in the one-dimensional case possible \cite{Veselic-96}.

We prove in Section \ref{sec4}  an approximation result (Theorem \ref{thmapp}) for the 
IDS of the multi-dimensional operator $ \H $, 
which enables us to prove Proposition \ref{techThm}.
In our case however periodic b.c.~seem to be more efficient than Dirichlet b.c.~since
$ \H $ is a perturbation of a periodic operator.

In \cite{Klopp-1999} it was proved that the IDS of $ \H $ exhibits  Lifshitz asymptotics
at a lower band edge $ \cal E $ if before the perturbation $ V_\omega $ the Floquet eigenvalues
of the periodic background operator $ H_0 $ at  $ \cal E $ were regular. Thus 
our approximation theorem can be applied to conclude localization.

%
%

\section{The Helffer-Sj\"ostrand formula: Functional\\ calculus with almost analytic functions\label{sec3}}

In this section we introduce the Helffer-Sj\"ostrand formula (\ref{HeSjFo}) 
which is exploited in Section \ref{sec4} to prove the IDS approximation result.

For an self-adjoint operator on $L^2(\RR^d) $ and a complex-valued measurable function
$ f : \RR \to \CC $ one can define the operator
\[
f(A)  \mbox{ with domain } D(f(A)) := \{ \psi \in L^2 (\RR^d) | \
f(A) \psi \in L^2 (\RR^d) \}
\]
via the spectral theorem. The latter is normally proved using Riesz'
representation theorem for $ C(K)^* $, where $K$ is a compact metric space,
and the Cayley-transform if $A$ is unbounded. Helffer and Sj\"ostrand \cite{HelfferS-89}
proved the following representation formula
\be
\label{HeSjFo}
f(A) := \frac{1}{2 \pi i} \int_\CC \frac{\partial \tilde{f} }{ \partial \bar{z} }
(z)  (z-A)^{-1} dz \wedge  d\tilde{z}
\ee
if $f$ is smooth and compactly supported. Here $ \tilde{f} : \CC \to \CC $
denotes an \begin{it} almost analytic extension \end{it} of $f : \RR \to \CC $.
Davies \cite{Davies-95c} uses equation (\ref{HeSjFo}) as a starting point to
develop systematically a functional calculus equivalent to the standard one.
For further details on the material of this section see his book.

\begin{dfn}
For  $ n \in \NN $ and $ f \in C^n_0 ( \RR, \CC ) $ define the almost analytic
extension (of order $n$) $ \tilde{f} : \CC \to \CC $ by
\be
\label{defext}
\tilde{f} (x,y) :=
\tilde{f}_n (x,y)  :=  \left( \sum_{r=0}^n f^{(r)} (x) 
\frac{(iy)^r}{ r!} \right) s(x,y)   \ ,
\ee
where we used the convention $ z := x + iy := (x,y) \in \CC $.
The cutoff function $ s $ is defined  with the abbreviation
$ \langle x \rangle := \sqrt{x^2+1} $ by the formula
\[
s(x,y) := t \left( \frac{y}{\langle x \rangle} \right), \ t \in C_0^\infty (\RR)
\]
with $ t(x) = 0 $ for $ |x| > 2$, $ t(x) = 1 $ for $ |x|< 1 $ and $ \|t'\|_\infty \le 2$.
\end{dfn}

With this choice of the almost analytic extension formula (\ref{HeSjFo})
holds true. If the support of $f$ is contained in $ [ -R , R ] $, $ \tilde{f}$
vanishes outside the set $ \{z \in \CC | \ x \in \supp f, |y| <2R +2\} $.
A calculation of the derivatives shows
\beq
\frac{ \partial \tilde{f}_n }{ \partial \bar{z} } (z)
& = & 
\frac{1}{2} \left( \frac{ \partial \tilde{f}_n }{ \partial x } +i \frac{ \partial \tilde{f}_n }{ \partial y } \right) (z)
\\
& = &
\nonumber 
  \frac{1}{2} f^{(n+1)} (x) \frac{ (iy)^n }{n!} \ s(x,y) + 
  \frac{1}{2} ( s_x (x,y) +i s_y(x,y) )\    \sum_{r=0}^n f^{(r)} (x) \frac{ (iy)^r }{r!}  \ .
\eeq
By calculating the partial derivatives of $s$ we see 
\be
|s_x + i s_y| \le \frac{6}{\langle x \rangle} \chi_{\{\langle x \rangle < |y|<2\langle x \rangle\}} \ ,
\ee
which shows that they vanish for $|y| \le 1$ since always $ \langle x \rangle = \sqrt{x^2+1} \ge 1 $.
Putting the bounds together we get
\beq
\label{DerExtEst}
\left| \frac{ \partial \tilde{f}_n }{ \partial \bar{z} } (x,y) \right|
& \le & 
\frac{1}{2n!} |f^{(n+1)}s| \ |y|^n + \frac{3}{\langle x \rangle} \chi_{\{\langle x \rangle < |y|<2\langle x \rangle\}}
\sum_{r=0}^n |f^{(r)} | \frac{ |y|^r }{r!}  \, .
\hspace*{\fill}
\eeq
Later on $f$ will be an approximation of the characteristic function
$ \chi_{[0,E  ]} $. It is going to have support inside $ [-E/2, 2E] $ and be
equal to $1$ on $[0,E ]$. One can choose $ f $ in such a way that $ \| f^{(n)} \|_\infty \le
C E^{-n} $ and
\be
\label{minderiv}
\NO f \ON_n := \sum^n_{i = 1} \| f^{(n)} \|_\infty \le \tilde{C} E^{-n}
\ee
for sufficiently small $E$. The constants $C,\tilde{C}$ are independent of $E$.

%
%
%

\section{IDS approximation theorem\label{sec4}} 

In this section we bound the difference of the IDS of the ergodic operator
$ H_\omega $ and its periodic approximation $ H_{\omega,l} $ which will be defined shortly. 
The estimate is contained in Theorem \ref{thmapp} which is the main technical result of this paper.
Furthermore, it enables us to show in Theorem \ref{LtforPA} that the IDS of the periodic approximation $ H_{\omega,l} $ exhibits a kind of Lifshitz tail, if the IDS N of the original operator $ H_\omega $ does so.
The periodic approximation $ H_{\omega,l} $ is defined by
\be
H_{\omega ,l}(x) := H_0(x) + \sum_{k \in \ZZ^d} \omega_{\tilde{k}} \:
u (x-k)
\ee
where $ \tilde{k} := k \, (\mbox{mod} (2l +1) \ZZ^d) $. For any $ l \in \NN $
and $ \omega \in \Omega $ it is a $ \mbox{(2l+1)}\ZZ^d$-periodic operator. Our
assumptions on $ u $ and $ \omega $ ensure that it is an infinitesimally small
perturbation of $ H_0 $, uniformly in $ l $ and $ \omega $. Hence it is a lower
bounded symmetric operator which is self-adjoint on the domain $ W^2_2 (\RR^d) $.
Its IDS is defined by (cf.~\cite{ReedS-78,Shubin-79,Sjoestrand-91})
\be
\label{DefIDSper}
N_{\omega,l}(E) := N(H_{\omega,l},E) := (2 \pi )^{-d}
\sum_{n \in \NN} \int_{B_l} \chi_{ \{ E_n(\theta) < E \} } \, d\theta \ .
\ee
Here $ E \in \RR $ is an energy value, $ E_n(\theta) $ is the $n$-th eigenvalue of
$ H_{\omega,l} |^\theta_{\Lambda_{2l + 1}} $ and 
\be
\label{BriZ}
 \theta \in B_l := \l [ \frac{- \pi}{2l +1}, \frac{ \pi}{2l +1} \r ]^d 
\ee
 if $ H_\omega $ is $ \ZZ^d$-ergodic. For some other
Euclidean lattice it has to be replaced by the basic cell of the corresponding dual lattice  $ \Gamma^* := \{ \gamma^* \in (\RR^d)^* = \RR^d | \, \forall \gamma \in \Gamma : \gamma^* \cdot \gamma \in 2 \pi \ZZ \} $.
We prove the following approximation result:

\begin{thm}
\label{thmapp}
Let $ H_\omega $ be defined as in Section \ref{sec1} and $ H_{\omega,l} $ as above. Denote
by $ N $, respectively $ N_{\omega,l} $ the corresponding IDS'. For a real valued
function $ g \in C^{n+1}_0 $ with support in $ [ -1/2, 1/2] $ we have
\bea
\left| \EE \left( \int_\RR g(x) dN_{\omega,l} (x) \right) -
\int_\RR g(x) dN(x) \right|
\le \const \ |\supp g |  \ \NO g \ON_{n+1}  \,  l^{-n+2d+1}
\eea
for sufficiently large $l \in \NN$.
\end{thm}
The proof is split into several lemmata. Remark \ref{remapp} and Lemma \ref{lemapp1}
are taken from Section 5.2 of \cite{Klopp-1999}.
We denote with $ \chi_l $ the characteristic function of the periodicity
cell  $ \Lambda_{2l + 1} := \{ x \in \RR^d | \ \| x \|_\infty \le l + 1/2 \} $
of $ H_{\omega,l} $ and by $ \chi_{l,\gamma} (x) := \chi_l (x-\gamma) $ its
translation by $ \gamma \in \ZZ^d $.

\begin{bem}
\label{remapp}
Note that one can infer from \cite{Carmona-84b,CarmonaL-90},\cite{PasturF-92} and \cite{Klopp-1999} the following equalities
\be
\int_\RR  g(x) \, dN(x) = \EE ( \Tr \chi_0  \, g(H_\omega) \chi_0 )
\ee
respectively
\be
\int_\RR  g(x) \, dN_{ \omega,l} (x) = ( 2l +1)^{-d}  ( \Tr \chi_l \,  g(H_{\omega ,l}) \chi_l ) \  .
\ee
Using the decomposition
\[
\chi_{\, l} = \sum_{ k \in \ZZ^d , |k| < 2l + 1} \chi_{\, 0,\, k}   \   , 
\]
the $ (2l+1) \ZZ^d$-periodicity of $ H_{\omega, l} $ and the i.i.d. property of 
$ ( \omega_k)_{k \in \ZZ^d} $ one gets
\be
\EE \l ( \int_\RR  g(x) \, dN_{\omega,l} (x) \r ) = \EE \l ( \Tr \chi_0  \,g(H_{\omega, l}) \chi_0 \r ) \ . 
\ee
Since $ \h $ is uniformly lower bounded there exists a $ \lambda \ge 0 $ such that 
$\, \Id \le \lambda + \h $ and $\, \Id \le \lambda + \H $ for all $ l, \omega $. 
From \cite{Simon-82c} we know that the operator  
\mbox{$ \chi_l (\lambda + \h )^{-q} (z-\h)^{-1} $} is trace class for all $ q > d/2 $. 
Using results from the appendix of \cite{Klopp-95d} we infer
\beq
\| \chi_{0,\beta } (z - \H )^{-1} (\lambda + \H )^{-q} \chi_0 \|_{\Tr} & \le & \frac {\tilde{C}_1} {|y|} \exp (-|y| \, |\beta| /{\tilde{C}}_1 )
\eeq
for some $ {\tilde{C}}_1 \ge 1$ independent of $ \omega $. 
This estimate is in fact a sophisticated version of the Combes-Thomas argument 
which we encountered already in Section \ref{sec2}. A simple resolvent estimate gives
\be
\label{PotDec}
\| \chi_{0 } (z - \h )^{-1}  T_\gamma \,  u \,  \chi_{0,\beta +\gamma} \|_{{\cal L}(L^2(\RR^d))} 
 \le  \frac {\tilde{C}_1} {|y|} \| \chi_{0,\beta +\gamma} u \|_{L^p}  \ ,
\ee
where $ T_\gamma $ is the translation by $ \gamma \in \ZZ^d $. 
As the single site potential $ u $ decays exponentially, inequality (\ref{PotDec})
gives a exponential bound in $ - | \gamma + \beta |$. If one assumes that $ u $ decays 
polynomially with a sufficiently negative exponent, one still can carry trough 
the proof of Theorem \ref{thmapp}.
\end{bem}

\begin{lem}
\label{lemapp1}
If $ g  \in C_0^{n+1} $ and $ \tilde{f} $ is an almost analytic extension of 
$f(x) := (\lambda + x )^q g(x)$, one has
\bea
\lefteqn{
\l | \EE \l ( \int_\RR g(x) \, dN_{ \omega,l} (x) \r ) -\int_\RR g(x) \, dN (x) \r |
}
\\ 
& \le &
\frac{C_1}{2 \pi}  \int_\CC |y|^{-2}  \l | \df (x,y) \r |
 \Biggl (
 \sum_{\beta \in \ZZ^d \atop {\gamma \in \ZZ^d,|\gamma | > l}}  
\| \chi_{0,\gamma + \beta} u \|_{L^p} \exp( -|y| \, |\beta| /C_1) \Biggr )  dx \, dy
\eea
\end{lem}

\begin{pro}
We use without explicit reference the equations collected in the above Remark \ref{remapp}
 and the Helffer-Sj\"ostrand formula (\ref{HeSjFo}). 
Let $ \NN \ni q > d/2 $. If we multiply
\be
g(\h) = \frac{i}{ 2 \pi} \int_\CC \df (z)  ( z - \h)^{-1} ( \lambda + \h)^{-q} \, dz \wedge d\bar{z}
\ee
by the characteristic function $ \chi_0 $ of $ \Lambda_{1} $ we get a trace-class operator and consequently
\be
\Tr ( \chi_0 \, g(\h) \chi_0) =  
\frac{i}{ 2 \pi} \int_\CC \df (z) 
\Tr  ( \chi_0 \, ( z - \h)^{-1} ( \lambda + \h)^{-q} \chi_0 )   \, dz \wedge d\bar{z} \ .
\ee
The same formula holds with $ \H $ substituted for $ \h $.
To bound the trace of $ \chi_0 \, ( \h - \H ) \chi_0 $ in mean we estimate
\bea
\| \chi_0 \,   ( z - \h)^{-1} ( \lambda + \h)^{-q} \chi_0 -
\chi_0 \,   ( z - \H)^{-1} ( \lambda + \H)^{-q} \chi_0       \|_{\Tr} \le \Sigma_1 + \Sigma_2
\eea
by the two summands
\bea
\Sigma_1 &  = &  \|  \chi_0  \l ( (z - \h )^{-1} - (z - \H )^{-1} \r ) ( \lambda + \H )^{-q} \chi_0 \|_{\Tr}
\\
& = & 
\bigg \|  \chi_0  \l ( (z - \h )^{-1} 
\l (  \sum_{ \gamma \in \ZZ^d, |\gamma |>l} (\omega_{\tilde{\gamma}} -\omega_\gamma ) u(x -\gamma) \r )
(z - \H )^{-1} \r )
\\
& & \times \
( \lambda + \H )^{-q} \chi_0    \bigg \|_{\Tr}
\eea
and
\bea
\Sigma_2 & = & \|  \chi_0   ( z - \h )^{-1} \l ( ( \lambda + \h )^{-q} - ( \lambda + \H )^{-q} \r ) \chi_0 \|_{\Tr}
\\
& = &
\bigg  \|  \chi_0   ( z - \h )^{-1} \: \sum^q_{m = 1} ( \lambda + \h)^{m-q-1} 
\l ( \sum_{\gamma \in \ZZ^d, |\gamma| > l} (\omega_{\tilde{\gamma}} -\omega_\gamma ) u(x -\gamma) \r )
\\
& & \times \
( \lambda + \H)^{-m} \chi_0 \bigg \|_{\Tr}  \  ,
\eea
where in the last equality we used an iterated resolvent formula.
Since $ | \omega_{\tilde{\gamma}} -\omega_\gamma | \le \omega_{\max} $ 
and by standard bounds for the trace norm $ \| \cdot \|_{\Tr} $ we have
\bea
\Sigma_1 
& \le & 
\omega_{\max} \sum_{ \beta \in \ZZ^d }  \sum_{ \gamma \in \ZZ^d, |\gamma |>l}
\l \| \chi_0 ( z - \h )^{-1}  u(x -\gamma)  \chi_{0,\beta} \r \|_{ {\cal L} ( L^2 ( \RR^d))}
\\
& & \times \
\| \chi_{0,\beta} (z- \H)^{-1} ( \lambda + \H)^{-q} \chi_0 \|_{\Tr}
\\ 
& \le &
\frac{C_1}{ |y|^2}     \sum_{ \beta \in \ZZ^d }  \sum_{ \gamma \in \ZZ^d, |\gamma |>l}
\| \chi_{0,\gamma + \beta} u \|_{L^p} \exp( -|y| \, |\beta| /C_1) 
\eea
As $ \Sigma_2 $ can be bounded in the same way, our lemma is proved.
\end{pro}
Up to now we followed the proof of Theorem 5.1 of \cite{Klopp-1999} almost literally. 
From now on we need sharper and more explicit estimates because later we will have to take the limit $ l \to \infty $ simultaneously with an approximation $ g \to \chi_{[0,E]} $. Special care is needed because the parameters $ E $ and $ l $ are functions of each other.

\begin{lem}
\label{lemapp2}
If we choose the constant $ C_2 $ sufficiently large and $ C_3 $ sufficiently small (depending only on $d,\delta_2,\delta_3$ and $C_1$), we have for all $ y $ with $ 0 \neq|y| \le 3 $:
\be
\label{doublesum}
  \sum_{\beta \in \ZZ^d}  \sum_{\gamma \in \ZZ^d \atop |\gamma | > l}
\| \chi_{0,\gamma + \beta} u \|_{L^p} \exp( -|y| \, |\beta| /C_1) 
 \le 
C_2 \, e^{ -C_3|y| l} \, |y|^{-2d} .
\ee
\end{lem}
The proof of this and the following lemma are given in the appendix.
\begin{lem}
\label{lemapp3}
Let $ f $ be in $  C^{n+1}([-1/2,1/2]) $ and $ \tilde{f} $ its  almost analytic extension of order $ n $. 
There exists a $ l_1 := l_1 (d,n,C_3) < \infty $ such that we have  for all $ l \ge l_1 $:
\bea
\int_\CC \l | \df (x,y) \r | |y|^{-2d-2} e^{-C_3 |y| l} \, dx \, dy
& \le &
2 C_3^{-n+2d+2} \NO f \ON_{n+1} \: |\supp f| \: l^{-n+2d+1}  \  .
\eea
\end{lem}

We have to bound the derivatives of $ f := (\lambda + \cdot )^q g $ in terms of the derivatives
of $ g $ itself.  A simple calculation using Leibniz' formula shows  $ \NO f \ON_{n+1} \le C_4 \,  \NO g \ON_{n+1} $,
 where $ C_4 $ depends only on $ n,q $ and $ \lambda $.

We collect the estimates of Lemma \ref{lemapp1}, \ref{lemapp2} and \ref{lemapp3} and  write down the needed inequalities for our difference of integrals with respect to $ N $ and $ N_{\omega,l} $.
\bea
\lefteqn{
 \ \left| \EE \left( \int g(x) dN_{\omega,l}(x) \right) -\int g(x) dN(x) \right|
} 
\\
& \le &
\frac{1}{2 \pi}
\int_{\CC} dx\, dy \ 
 \frac{C_1}{|y|^2} 
\left| 
\frac{ \partial \tilde{f} }{ \partial \bar{z} } (x,y) \right|
\left (
\sum_{ \beta \in \Gamma } \sum_{ \gamma \in \Gamma,  \atop | \gamma | > l  }
\| \chi_{0, \beta + \gamma } u \|_{ L^p  } \exp (- |y|  \, | \beta |  / C_1  )
\right )
\\ 
 & \le &
\frac{1}{2 \pi}
\int_{\CC} dx\, dy \ 
\delta_2 C_1 
\left| 
\frac{ \partial \tilde{f} }{ \partial \bar{z} } (x,y) \right|
C_2 |y|^{-2d-2} \exp(- C_3 |y| l)
\\
& \le &
\frac{\delta_2 C_1 C_2 }{ \pi  C_3^{n-2d-2} }
   \ | \supp f |   \    \NO f \ON_{n+1} l^{-n+2d +1}
\\
& \le &
C_5   \ | \supp f |   \    \NO g \ON_{n+1} \, l^{-n+2d +1}
\eea
if we choose $l\ge l_1 $ and set 
$ C_5 := \frac{ \delta_2 C_1 C_2  C_4}{ \pi  C_3^{n-2d-2} } $. 
This proves Theorem \ref{thmapp} with $ C_5 $ as the constant on the rightern side.
\par \hspace*{\fill} {\bfseries q.e.d.} \\[1ex] 
   
The IDS approximation result (Theorem \ref{thmapp}) gives information about $N_{\omega,l}$ if properties of $N$ are known. Exploiting this fact, we want to show that $N_{\omega,l}$ is ''small'' in the energy region where $N$ exhibits a Lifshitz tail.
To this end take $ g \in C_0^{n+1}(\RR, [0,1]) $ with $ g (x) = 1 $ for all $x \in [0,E]$ and
support in $ [-E/2 , 2E] $. Moreover let $g$ have minimal derivative in the sense of inequality (\ref{minderiv}).
We estimate
    \begin{multline}
    \EE [ N_{ \omega , l} (E) - N_{ \omega , l} (0) ]
    \leq  
    \EE \biggl (  \int g \  dN_{ \omega ,l }  \biggr ) 
    \\ 
    \leq 
    \int g  \,  dN + \biggl | \EE \biggl (\int  g \  dN_{ \omega , l} \biggr ) - \int  g \  dN \biggr|   
    \ .
    \end{multline}
Let $E/2$ be smaller than the gap width $b'$ below the spectral band edge $0$. Since $\supp N = \sigma(H_\omega)$ a.s. (c.f.\cite{PasturF-92}) it follows for $ l \ge l_1 $
\be
\label{NE-N0}
\EE [ N_{ \omega , l} (E) - N_{ \omega , l} (0) ]
\le  
  N(2E) -N(0) \, + \, C_6 \, E^{-n} \, l^{-n +2d+ 1}  \ ,
\ee
where we used Theorem \ref{thmapp} and equation (\ref{minderiv}). 
If $ N $ has Lifshitz
asymptotics at the lower band edge $ 0 $, as defined in equation (\ref{LifTaildef}), there
exists an energy value $ E_1 $ such that
\be
\label{useLifTail}
N(E) - N(0) \le \exp (-E^{-d/4} ) \ \ \forall  E \in [ 0, E_1] \ .
\ee
Together with (\ref{NE-N0}) this gives
\be
\EE [ N_{ \omega , l} (E) - N_{ \omega , l} (0) ]
\le  
e^{-(2E)^{-d/4} } + \, C_6 \, E^{-n} \, l^{-n +2d+ 1}   \ \forall  E \in [ 0, E_1 /2] \ .
\ee
For  $ \alpha \in ]0,1[ $ we set $ E := 2 l^{-\alpha} $ . This implies
    \begin{eqnarray}
    \label{IDSpolSmall}
    \EE [ N_{ \omega , l} (E) - N_{ \omega , l} (0) ]
    & \le & 
    \exp (-(4l^{-\alpha})^{-d/4} ) + \, C_6 \, (2l^{-\alpha})^{-n} \, l^{-n +2d+ 1}
    \nn
    \\
    & = & 
    \exp (-4^{-d/4}\,  l^{\alpha d/4}) + \, C_6 2^{\alpha n} \, l^{-n(1-\alpha) +2d+1}
    \nn 
    \\
    & \le &
    C_6 2^n \, l^{-n(1-\alpha) +2d+1}
\end{eqnarray}
if $ l \ge l_2 := l_2 (d,n, \alpha, C_6, b', E_1 ) $. Thus we have proven that the Lifshitz tail
of $N$ implies a similar asymptotic behaviour of the IDS  of the periodic approximation $H_{\omega,l}$  as stated in the following
\begin{thm}
\label{LtforPA}
Let $ N $ and $ N_{\omega, l} $ be the IDS of $ \H $ and $ \h $ respectively, $n \in \NN$ and $ \alpha \in ]0,1[ $. 
If $ N $ has a Lifshitz tail at the lower band edge $ 0 $, there exist a $C_7 < \infty$ such that
\be
\EE [ N_{ \omega , l} (2l^{-\alpha}) - N_{ \omega , l} (0) ] \le
 C_7 \,  l^{-n(1-\alpha) +2d+1}
\ee
for  sufficiently large $ l $.
\end{thm}

%
%
%
%
%
\section{Sparsity of states near the lower band edge\label{sec5}}
We want to estimate the probability of finding an eigenvalue of  $ \h (\theta) $
in a small energy interval $ I \ni 0 $, assuming that $N$ exhibits a Lifshitz tail at $0$. 
Here $ \h (\theta) := \h |^\theta_{\Lambda_{2l + 1}} =  \H |^\theta_{\Lambda_{2l + 1}} $
is the operator $ \h $ restricted to $ L^2 (\Lambda_{2l + 1}) $ with $\theta$-boundary
conditions. The following lemma allows to bound this probability using the IDS of $ \h $.
\begin{lem}
\label{iniscalLem1}
\bea
\label{exsred}
\int_{ \theta \in B_l } \! d \theta \,  
\PP  ( \{ \omega | \ \sigma ( \h  (\theta)) \cap [0, E[ \, \neq \emptyset \} )      
\leq 
( 2 \pi )^d \, \EE (  N_{\omega,l} (E) -  N_{\omega,l} (0) ) \ .
\eea
\end{lem}
\begin{pro}
\bea
& 
\int_{ \theta \in B_l } \, d \theta \,  \PP  ( \{ \omega | \ \sigma ( \h  (\theta)) \cap [0, E[  \, \neq \emptyset \} )
&
\\
\\
 \leq &
| \Lambda_{2l + 1} |  \,
\int_{ \theta \in B_l } \, d \theta \,  \EE ( N( \h  (\theta),E) -  N( \h  (\theta),0)) 
& \mbox{\footnotesize \v{C}eby\v{s}ev inequality }
\\
\\
 = &
| \Lambda_{2l + 1} |  \,
\EE ( \int_{ \theta \in B_l } \, d \theta \,  
( N( \h  (\theta),E) -  N( \h  (\theta),0))
&  \mbox{\footnotesize Fubini's theorem }
\\
\\
 = &
( 2 \pi )^d \
\EE (  N_{\omega,l} (E) -  N_{\omega,l} (0) ) 
& \mbox{\footnotesize  equations (\ref{DefIDS},\ref{DefIDSper})}
\eea
\end{pro}
Since the MSA works with specific boundary conditions, e.g. periodic ones, we have
to get rid of the average over $ \theta \in B_l $ in the last bound.
This is possible using the Lipschitz-continuity  in $ \theta $ of the eigenvalues of $ \h (\theta) $.

\begin{lem}
\label{iniscalLem2}
For any fixed $ \theta_0 \in B_l $ and  $ E < 1$  we have
\be
\label{H1}
\PP ( \{ \omega | \, \sigma ( H_{ \omega,l } ( \theta_0 )) \cap [ 0 , E [ \, \neq \emptyset \} ) 
  \leq 
\frac{ (2 \pi)^d }{  | B_l |}
\EE( N_{ \omega,l } (E + C_9 l^{-1} ) - N_{\omega,l} (0) )   \ .
\ee
\end{lem}
\begin{pro}
The eigenvalues of $ H_{ \omega, l} ( \theta ) $ are  Lipschitz continuous in $ \theta $, 
so we have :
\bea 
| E_j (  H_{ \omega, l} ( \theta ) ) - E_j(  H_{ \omega, l} ( \theta' ) ) | 
\leq \Xi_{j,l} | \theta - \theta'| 
\eea
for some $ \Xi_{j,l}  > 0 $. 
One can choose the  $ \Xi_{j,l} $ independent of $ j $ and $l $ only as
a function of $ E_j(\h (\theta) ) $. As we consider only eigenvalues in the energy interval
$ [0, E[ \subset [0,1[ $ even this dependence can be eliminated. Thus we can find 
$ \Xi > 0 $ such that
\bea
\Xi  \ge \Xi_{j,l}  \ \forall l,j \  .
\eea
Now we can estimate :
\begin{multline}
\label{Lip+av}
\PP ( \{ \omega |  \ \sigma (  H_{ \omega, l} ( \theta_0 ) ) \cap [ 0,E[ \, \neq \emptyset \} )
 =  
\PP ( \{ \omega |  \ \exists j \in \NN : \ E_j (  H_{ \omega, l} ( \theta_0 ) ) \in [ 0,E[  \ \} )
\\
 = 
 \int_{ \theta \in B_l } \frac{d \theta}{| B_l |} \,  
\PP ( \{ \omega |  \ \exists j \in \NN : \ E_j (  H_{ \omega, l} ( \theta_0 ) ) \in [ 0,E[  \ \} )
\end{multline}
If $ E_j (  H_{ \omega, l} ( \theta_0 ) ) \in [ 0,E[  $ then $
E_j (  H_{ \omega, l} ( \theta ) ) \in [ 0,E +  \Xi \diam (  B_l  )[  \ \
\forall \theta \in B_l $. Using $\diam ( B_l ) \le C_8 \ l^{-1}$ we bound (\ref{Lip+av}) by
\bea
\lefteqn{
 \int_{ \theta \in B_l } \frac{d \theta}{| B_l |} \,  
\PP ( \{ \omega |  \ \exists j \in \NN : \ E_j (  H_{ \omega, l} ( \theta ) ) \in [ 0,E + C_9 l^{-1} [  \ \} )
}
\\
& = & 
 \int_{ \theta \in B_l } \frac{d \theta}{| B_l |} \, 
 \PP ( \{ \omega |  \ \sigma(  H_{ \omega, l} ( \theta ) ) \cap [ 0,E + C_9 l^{-1} [  \neq \emptyset \ \} )
\\
& \leq & 
\  (2 \pi)^d   \  | B_l |^{-1}\  \EE( N_{ \omega , l}( E + C_9 l^{-1} ) - N_{\omega,l}(0) ) 
\eea
\end{pro}
We choose now $ 0 < \alpha < 1$ and $ E := l^{-\alpha} $ similarly as before.
Thus for $ l \ge l_3 $ the bound
$
E + C_9 \, l^{-1} \le  2 l^{-\alpha}
$
is valid, with $ l_3 $ depending on $ \alpha $ and $C_9$.
As the IDS is monotone increasing in the energy, this implies
\bea
N_{\omega,l} ( E+ C_9 \, l^{-1} ) \le N_{\omega,l} ( 2 l^{-\alpha} )  \  .
\eea
If $ N $ has Lifshitz tails, we estimate as in Theorem \ref{LtforPA}:
\bea
\EE ( N_{\omega,l} ( 2 l^{-\alpha} ) - N_{\omega,l} (0) ) 
\le C_7 \, l^{ -n (1 - \alpha) +2d +1}
\eea 
for $ l \ge l_2 $.
In this way we obtain from Lemma \ref{iniscalLem2}
\be 
\label{ProbPolSmall}
\PP (\{ \omega| \: \sigma( \h (\theta_0) ) \cap [0,l^{-\alpha}[ \, \neq \emptyset \})
 \le 
C_{10} l^{-n ( 1-\alpha) +3d +1}
\ee
since $ |B_l|^{-1} \le const \,  l^d $ where the constant depends only on the dimension.
The probability in (\ref{ProbPolSmall}) can be bounded by $ l^{-q} $ for arbitrary $ q > 0 $ if
\beq
-n (1-\alpha) +3d +1 & <  & -q
\nn
\\
\label{condAlphN}
\aeq
n (1-\alpha)         & >  & q   +3d +1  
\eeq
and  $ l \ge l_4:=l_4 (d,n,\alpha,q, C_{10})  $ is sufficiently large.
It is obvious that for any $ 0 < \alpha < 1 $ we can choose $ n $ in such a way that the relation
(\ref{condAlphN}) is valid.  

Similarly, for any fixed $ n > q +3d +1 $ it is possible 
to choose $ \alpha $ sufficiently small, so that (\ref{condAlphN}) holds. Particularly we can choose $\alpha$ from $]0,1/4[$.\\[0.2em]

Recall that if $H_0$ has regular Floquet eigenvalues at the lower spectral band edge $ 0 $,
the IDS $N$ of $ \H := H_0 + V_\omega $ exhibits Lifshitz asymptotics at $0$. 
Thus we proved Proposition \ref{techThm} with $ l_0 := \max_{i = 1}^4  l_i $.

\section{Appendix}
{\bf Proof of Lemma \ref{lemapp2}.}
By comparing the Euclidean and sup-norm, the sum in (\ref{doublesum}) can be bounded by a constant times the integral
\be
\label{app1}
\int_{\RR^d} dx \int_{\|\xi\|_2 > l} d\xi \, e^{-\delta_3\kappa \|x+\xi\|_2} e^{-|y|\kappa \|x\|_2/C_1}, 
\qquad \kappa := d^{-1/2}.
\ee
Substituting $ x' = (|y| \delta_3\kappa/6C_1) (2x+\xi), \xi' = (|y| \delta_3\kappa/6C_1) \xi$, using the parallelogram identity for $\|\cdot\|_2$ and $|y| \le 3, C_1 \ge 1$ we estimate (\ref{app1}) by 
\bea
2^{d} \left( \frac{3C_1}{\delta_3\kappa|y|}\right)^{2d} 
\int\limits_{\RR^d} dx' \hspace{-0.4cm} \int\limits_{\|\xi'\|_2 >\delta_3 \kappa |y|l /6C_1} \hspace{-0.8cm} 
d\xi' \, e^{- \|x'\|_2- \|\xi'\|_2}
\le 
const \, |y|^{-2d} \, \exp \left( - \frac{\delta_3\kappa}{12C_1} |y|l \right)
\eea
where the constant depends only on  $d,\delta_3$ and $C_1$. 
\par \hspace*{\fill} {\bfseries q.e.d.} \\[1ex]
{\bf Proof of Lemma \ref{lemapp3}.}
We use inequality (\ref{DerExtEst}) and consider first the term:
\be
\label{app2}
\int_{\CC} dx \, dy 
|y|^{-2d-2} \exp(-C_3 |y| l)
             \frac{3}{\langle x \rangle} \chi_{\{\langle x \rangle < |y|<2\langle x \rangle\}}
\sum_{r=0}^n |f^{(r)}(x) | \frac{ |y|^r }{r!} \ .
\ee
The properties of $\langle x \rangle$, $s$ and $f$ ensure $\langle x \rangle \ge 1, 1<|y|<3 $, thus 
\bea
 (\ref{app2}) \le  6\int_{ \supp f} \!\!\!\!\!  dx \int_{[1,3]} dy \  e^{-C_3  l}  \sum_{r=0}^n |f^{(r)} | \frac{ 3^r }{r!}  
\le  60  \ | \supp f | \ \NO f \ON_n \, e^{-C_3  l}.
\eea
Now we turn our attention to the other summand in (\ref{DerExtEst})
\begin{multline}
\int_{\supp f} \!\!\!\!\!\! dx \int dy \  |y|^{n-2d-2} e^{-C_3  |y| l}   \,  \frac{|f^{(n+1)} (x) |}{2n!} 
\label{app3}
\\
\le 
C_3^{-n+2d+2} \NO f \ON_{n+1}  \  |\supp f |  \  l^{-n+2d+1}  \  .
\end{multline} 
For  sufficiently large $ l $, i.e.  $ l \ge l_1'(d,n,C_3) $, we have
\bea
(\ref{app2}) + (\ref{app3})
\le 
2 C_3^{-n+2d+2} \NO f \ON_{n+1} \: |\supp f|  \: l^{-n+2d+1} .
\eea
\par \hspace*{\fill} {\bfseries q.e.d.} \\[1ex]
 
\def\cprime{$'$}

\end{document}